# TITLE MASERS (Entry Type: Long Entry)[1]

Elizabeth Humphreys

European Southern Observatory

Karl-Schwarzschild Strasse 2

Garching bei Munchen

Germany

Email: ehumphre@eso.org

## Definition

An astrophysical MASER (Microwave Amplification by Stimulated Emission of Radiation) is a source of stimulated spectral line emission. Maser emission is observed from the circumstellar envelopes of evolved stars, molecular clouds/star-forming regions, active galactic nuclei, supernova remnants, comets, and the Saturnian moons. It arises from molecules such as water ($H_2O$), hydroxyl radicals (OH), methanol ($CH_3OH$), formaldehyde ($CH_2O$), silicon monoxide (SiO), ammonia ($NH_3$), silicon sulphide (SiS), hydrogen cyanide (HCN), and from atomic hydrogen recombination lines. Masers are compact, of high brightness temperature, and often display narrow spectral line widths, polarized emission and variability. Free electron-cyclotron astrophysical masers additionally exist.

## History

Astrophysical maser, or microwave laser, emission was first observed in the 1960s [1]. Initially, the nature of the emission mechanism was not understood and it was hypothesized that a new element, labelled 'mysterium', could be responsible. However, the spectral line was rapidly identified as a transition of the OH (hydroxyl) radical [2], and it was soon realised that the signal characteristics could be explained in terms of amplification by stimulated emission of radiation [e.g., 3,4].

## Overview

Astrophysical masers are regions of gas in which the amplification of radio or microwave wavelength radiation takes place. Maser amplification occurs at a wavelength corresponding to the energy difference between two energy states of a molecular/atomic species in the gas. In order for net stimulated emission to occur, a higher number of molecules must exist in the upper energy state than in the lower state. The populations of the levels are then said to be *inverted*, and this is a pre-requisite for maser action. The energy source that maintains the population inversion is known as the maser pump, and can be radiative, collisional or chemical in nature

---

[1] Word range of Entry: approx. 1,650 words excl. references, as agreed with the Editors-in-Chief

(or a combination of these processes). The spontaneous or continuum radiation that stimulates maser action is amplified exponentially as a function of distance through the gas as long as the pump rate exceeds that of the maser stimulated emission rate. This is the *unsaturated* regime. As amplification continues, maser radiation may become sufficiently strong for the maser stimulated emission rate to rival that of the pump. The maser is then said to start to *saturate* and amplification is linear as a function of distance. Maser radiation is partially-coherent, highly beamed and can display high degrees of polarization [5,6,7,8].

## Basic Methodology

Observationally, how is it possible to determine whether maser emission is being observed? One way is to measure the *brightness temperature* of the emission. If the brightness temperature exceeds the temperature at which molecules can exist, a few thousand Kelvin, then the emission cannot have come from a thermally –excited source. Masers often have brightness temperatures of $> 10^9$ K, and are compact compared with the scales of the astrophysical objects with which they are associated (e.g., stars, galactic nuclei). For example masers have sizes of order $10^7$ m in comets, and masers in galactic cores can have scale sizes up to $10^{19}$ m [6]. Other observational features that indicate maser emission are narrow, sub-thermal line widths in spectra and time variability. Lines from different transitions of the same molecular species may also display ratios inconsistent with thermal emission. Maser polarization is typically interpreted as due to the Zeeman effect in the presence of magnetic fields.

Maser emission often involves molecular rotational transitions, with the masers typically pumped in warm, dense gas. For example, the best-studied SiO masers are those at 43 and 86 GHz ($\lambda$=7 and 3 mm respectively) originating from the v=1 and 2 J=1 – 0 and 2 -1 transitions. They are pumped at gas densities of $n(H_2)=10^{10\pm1}$ cm-3 and kinetic temperatures of $T_k \geq 1500$ K [e.g, 5]. The best-studied water masers are those at 22 GHz ($\lambda$=1.3cm) originating from the $6_{16} - 5_{23}$ transition. They are pumped in gas of $n(H2)=10^8$-$10^{10}$ cm$^{-3}$ and $T_k = 400 - 1000$ K [e.g., 5].

When astrophysical objects associated with masers are imaged at high angular resolution, such as at sub-milli-arcsecond resolution using Very Long Baseline Interferometry (VLBI), multiple maser features are often observed toward the same object. In addition, as masers from multiple transitions of the same molecular species can amplify along the same regions of gas, masers at different frequencies are often detected toward the same source. Finally, at least in the case of evolved stars and star-forming regions, masers from different molecular species are also often found toward the same object.

## Key Research Findings

The circumstellar envelopes of oxygen-rich evolved low-mass and Supergiant stars commonly harbour SiO, $H_2O$ and OH masers. Radio VLBI observations indicate that SiO masers occur close to the star, typically within about 4 $R^*$ [9]. SiO maser emission has been detected from v=0 to 4, up to at least J=8-7 ($\nu$=344 GHz; $\lambda\approx$1 mm) in evolved stars and SiO isotopologue masers are also observed [e.g., 10]. $H_2O$ masers at 22 GHz form further from the star at a radial distance of about 10 $R^*$, with several other $H_2O$ maser transitions also detected at frequencies up to 658 GHz. Main line (1665 and 1667 MHz) OH masers occur at similar radii to the 22 GHz $H_2O$ masers, but are believed to probe significantly less dense gas [6]. At much larger radii of typically 100 $R^*$, 1612 MHz OH masers are observed. Polarized emission of the three molecular maser species has been interpreted as circumstellar magnetic field measurements [11]. Proper motion studies of the masers have been use to trace gas dynamics in the circumstellar envelope, with both outflow and infall detected in the SiO maser zone due to stellar pulsation. Towards the less common carbon-rich stars, HCN, CS and SiS masers have been observed.

Star-forming regions in molecular clouds support a variety of maser species, with methanol, water and OH masers the most frequently detected. Maser emission is observed from low and high mass forming stars, however stellar masses significantly > 1 $M_{sun}$ are favoured [6]. Water masers at 22 GHz are often seen in protostellar outflows associated with shocked gas, and are more rarely thought to be associated with accretion disks. Other water masers, including those at frequencies of 183, 321 and 325 GHz, have also been detected. Methanol masers in star-forming regions are divided into two types, Class I and Class II, depending on the transitions present. Class I methanol masers are primarily collisionally pumped, and may trace outflow interactions with dense molecular gas. Class II methanol masers tend to be associated with HII regions, with the best-studied transitions occurring at frequencies of 6.7 and 12.2. GHz. Methanol, OH (mainly 1665 and 1667 MHz lines) and $H_2O$ masers are often present in the same source, although water and methanol masers usually appear in different regions. Again, maser polarization is used to determine magnetic field characteristics, with fields as high as 40 mG seen in OH masers. Rarer masers in star-forming regions include ammonia, formaldehyde, and SiO (currently detected in 3 high-mass star-forming regions only) [12].

Supernova Remnant (SNR)/Molecular Cloud interactions also give rise to maser emission. OH 1720 MHz masers have been detected from ~20 galactic SNRs, i.e., 10% of the galactic sources [13]. Methanol maser emission at 95 GHz has also been detected towards one SNR. Hot stars MWC 349A and Eta Carina harbour H-atom recombination masers over a range of frequencies [see e.g., 6].

Extragalactic maser emission, in some cases referred to as *megamaser* emission, is most commonly detected from 22 GHz $H_2O$ and OH 1665 and 1667 MHz lines, although formaldehyde maser emission at 6 cm is also known. OH megamaser emission occurs from nuclear starburst activity, on scales of ~100 pc, whereas 22

GHz water maser emission is observed on (sub-)parsec scales and traces the central engines of Active Galactic Nuclei (AGN) [14]. Over a hundred extragalactic water maser sources are now known, including a z=2.64 detection magnified by gravitational lensing [15]. Detections of extragalactic methanol masers are limited to the Large Magellanic Cloud and appear to be similar to Galactic methanol masers.

In the Solar System, masers have been detected from cometary comas. Cometary OH masers have a UV pumping mechanism and have been observed from at least sixteen comets, including Comet Hale-Bopp. Water masers at 22 GHz have also likely been detected. Water maser emission was detected after the impact of Comet Shoemaker-Levy 9 with Jupiter, and has also been found from the Saturnian moons Hyperion, Titan, Enceladus, and Atlas [16]. Molecular laser features from $CO_2$ at 10.33 μm were detected in the atmospheres of Venus and Mars. Searches have also been performed for molecular maser emission associated with exoplanets, but to date no emission sources have been reported.

## Applications

Masers are used as tools to probe the physical conditions, kinematics and magnetic fields of astrophysical sources at high angular resolution. Maser line ratios constrain radiative transfer models to determine gas temperatures and densities. Proper motion measurements enable tracing of gas kinematics. Maser polarization, when attributed to the Zeeman effect, can be used to determine magnetic field strength and morphology characteristics. Masers can be used to measure dynamical masses and to provide distance estimates. In the case of water masers in AGN, this has led to their use for performance of high-accuracy estimation of the Hubble constant, which may lead to constraint on the nature of Dark Energy [17,18]. OH masers are used to search for variations in the fundamental "constants" over cosmological time [19]. Galactic trigonometric parallax measurements will trace the number and location of the spiral arms in the Milky Way [20].

## Future Directions

New instruments will revolutionize the use of masers as diagnostic probes. The Atacama Large Millimetre/Submillimetre Array (ALMA) will perform imaging up to ~950 GHz, enabling observation of high frequency maser lines at angular resolutions of up to ~10 milli-arcseconds. At low frequencies (foreseen 0.1 – 25 GHz), the Square Kilometer Array (SKA) will provide very high sensitivity, allowing for searches for distant masers in the early Universe.

## References and Further Reading